%% file: llmlf.tex
\documentclass[sigconf, nonacm]{acmart}

\newcommand\vldbyear{2024}
\newcommand\vldbworkshop{The 1st International Workshop on Data-driven AI (DATAI)}
\newcommand\vldbauthors{\authors}
\newcommand\vldbtitle{\shorttitle} 
\newcommand\vldbavailabilityurl{}
\newcommand\vldbpagestyle{empty}

\usepackage{url}
\usepackage{verbatim}
\usepackage{graphicx}
\usepackage{multirow}
\usepackage{balance}
\usepackage{lscape}
\usepackage{float}
\usepackage{xcolor}
\usepackage{paralist}
\usepackage{booktabs}
\usepackage{xspace}
\usepackage{listings}
\usepackage{lipsum}
\usepackage{subcaption}
\definecolor{ForestGreen}{rgb}{0.0, 0.66, 0.47}
\definecolor{RubineRed}{rgb}{1.0, 0.0, 0.31}

\usepackage{dsfont}

\begin{document}
\title{LLM-assisted Labeling Function Generation for Semantic Type Detection}

\author{Chenjie Li}
\affiliation{%
	\institution{Illinois Institute of Technology}
	\country{United States}
}
\email{cli112@hawk.iit.edu}

\author{Dan Zhang}
\affiliation{%
	\institution{Megagon Labs}
	\country{United States}
}
\email{dan_z@megagon.ai}

\author{Jin Wang}
\affiliation{%
	\institution{Megagon Labs}
	\country{United States}
}
\email{jin@megagon.ai}

\input{src/sec0-abstract}
\maketitle
\setcounter{page}{1}

\pagestyle{\vldbpagestyle}
\begingroup\small\noindent\raggedright\textbf{VLDB Workshop Reference Format:}\\
\vldbauthors. \vldbtitle. VLDB \vldbyear\ Workshop: \vldbworkshop.\\ %\vldbvolume(\vldbissue): \vldbpages, \vldbyear.\\
%\href{https://doi.org/\vldbdoi}{doi:\vldbdoi}
\endgroup
\begingroup
\renewcommand\thefootnote{}\footnote{\noindent
This work is licensed under the Creative Commons BY-NC-ND 4.0 International License. Visit \url{https://creativecommons.org/licenses/by-nc-nd/4.0/} to view a copy of this license. For any use beyond those covered by this license, obtain permission by emailing \href{mailto:info@vldb.org}{info@vldb.org}. Copyright is held by the owner/author(s). Publication rights licensed to the VLDB Endowment. \\
\raggedright Proceedings of the VLDB Endowment. %, Vol. \vldbvolume, No. \vldbissue\ %
ISSN 2150-8097. \\
%\href{https://doi.org/\vldbdoi}{doi:\vldbdoi} \\
}\addtocounter{footnote}{-1}\endgroup

\ifdefempty{\vldbavailabilityurl}{}{
\vspace{.3cm}
\begingroup\small\noindent\raggedright\textbf{VLDB Workshop Artifact Availability:}\\
The source code, data, and/or other artifacts have been made available at \url{\vldbavailabilityurl}.
\endgroup
}

\input{src/sec1-intro}
\input{src/sec2-method}

\input{src/sec3-exp}

\input{src/sec4-conclusion}

\bibliographystyle{ACM-Reference-Format}
\bibliography{llmlf}

\end{document}

%% file: src/sec0-abstract.tex
\begin{abstract}
	Detecting semantic types of columns in data lake tables is an important application. A key bottleneck in semantic type detection is the availability of human annotation due to the inherent complexity of data lakes. In this paper, we propose using programmatic weak supervision to assist in annotating the training data for semantic type detection by leveraging labeling functions. One challenge in this process is the difficulty of manually writing labeling functions due to the large volume and low quality of the data lake table datasets. To address this issue, we explore employing Large Language Models (LLMs) for labeling function generation and introduce several prompt engineering strategies for this purpose. We conduct experiments on real-world web table datasets. 
	Based on the initial results, we perform extensive analysis and provide empirical insights and future directions for researchers in this field.
\end{abstract}

%% file: src/sec1-intro.tex
\section{Introduction}\label{sec-intro}

% para 1: brief overview of STD problem and its applications
Semantic type detection is an important task in many data preparation applications, such as data cleaning, schema matching, entity resolution and data discovery~\cite{DBLP:journals/pvldb/ZhangSLHDT20,DBLP:conf/sigmod/SuharaL0ZDCT22,DBLP:conf/sigmod/Wang0HK22,DBLP:conf/deem/Wang022}.
Given a table and a set of semantic labels, semantic type detection aims at identifying a type label for each column in the table so that each cell in the column has the same semantic types.
This task has attracted significant attention from the database community, and many solutions based on deep learning techniques, especially pre-trained Language Models (PLMs)~\cite{DBLP:journals/pvldb/ZhangSLHDT20,DBLP:conf/sigmod/SuharaL0ZDCT22,DBLP:journals/pacmmod/MiaoW23}, have been developed to improve overall performance.

% para 2: chanllenge in labeling resource, introduce data programming
Although such PLM-based solutions are effective, they have a high requirement of labeled training instances to perform fine-tuning.
Due to the large scale and complex structure of data lake tables, it is rather challenging to acquire high-quality human annotation for semantic type detection~\cite{DBLP:conf/sigmod/Fan00M23}.
We argue that a weak supervision approach, such as data programming~\cite{DBLP:conf/nips/RatnerSWSR16}, is a good solution to reduce the burdens of training data annotation.
In the data programming paradigm, users are asked to design label functions (LF) that provide labels to a subset of data at a much lower cost rather than manually label instances one by one.
Then a label model is learned to denoise and aggregate the weak labels from each LF.
Finally, the label model could predict labels over unlabeled corpus to provide annotated training data. 

% para 3: previous work of data programming (summarize related work), challenge in std task
Over the past decade, significant efforts have been made in the field of data programming.
\textsf{Snorkel}~\cite{DBLP:journals/pvldb/RatnerBEFWR17} proposed a probabilistic model to aggregate the user-written LFs.
\textsf{Snuba}~\cite{DBLP:journals/pvldb/VarmaR18} aimed at proposing explainable LF while \textsf{Nemo}~\cite{DBLP:journals/pvldb/HsiehZR22} and \textsf{WITAN}~\cite{DBLP:journals/pvldb/DenhamLSN22} focused on the problem of interactive data programming.
The recent advances in the era of Large Language Model (LLM), such as GPT-4~\cite{DBLP:journals/corr/abs-2303-08774} and LLaMA~\cite{DBLP:journals/corr/abs-2302-13971}, have shown powerful capability in various tasks in different fields.
Some recent efforts~\cite{DBLP:journals/corr/abs-2311-00739,DBLP:conf/acl/ZhangYSSZ22} have been made to harness LLMs for automate the generation of LFs for NLP tasks.
However, it is non-trivial to extend them to support the task of semantic type detection.
Compared with the tasks supported in the previous studies, semantic type detection usually has a much larger labeling space and cardinality of datasets.
For example, the number of semantic labels in the Gittable~\cite{DBLP:journals/pacmmod/HulsebosDG23} and TURL~\cite{DBLP:journals/pvldb/DengSL0020} corpus is 835 and 255, respectively. 
Meanwhile, the task with the largest number of labels in the WRENCH benchmarking~\cite{DBLP:conf/nips/ZhangYNWYYR21} only has 18 class labels.
The large number of class labels brings two extra challenges: on the one hand, it is rather difficult for users to manually write enough LF for each class; on the other hand, it brings new challenges for the scalability of label model such as \textsf{Snorkel} to handle such large number of LFs and seeding instances for weak supervision.

% para 4: our contribution
In this paper, we propose an end-to-end framework to conduct weakly supervision to generate LFs for semantic type detection with the help of LLM.
We systematically explore the strategies to construct LLM prompt for generating LFs given the seed instances of each label class.
Specifically, we find that it is essential to include both the contents and the ground truth label of the seed instance in the prompt so as to provide sufficient context for LLM to produce effective LFs.
To improve the scalability of \textsf{Snorkel} for semantic type detection, we develop a stacked label model to split the label space and allow the sub-models to run in parallel. 
We conduct experiments on widely-used tabular datasets and evaluate both the quality of the generated LFs and the effect of training end models with the datasets annotated by such LFs.
Finally, we make an in-depth analysis of the preliminary results and provide some directions for the future work.

%% file: src/sec2-method.tex
\section{Methodology}\label{sec-method}

\subsection{Overview}\label{subsec-overall}
% \begin{figure*}[h]
% 	\centering
% 	\includegraphics[width=0.9\textwidth]{fig/overall}
% 	\vspace{-1em}
% 	\caption{Overall Workflow of Labeling Pipeline with LLM}
% 	\label{fig:system}
% \end{figure*}

\begin{figure}[h]
	\centering
	\includegraphics[width=0.45\textwidth]{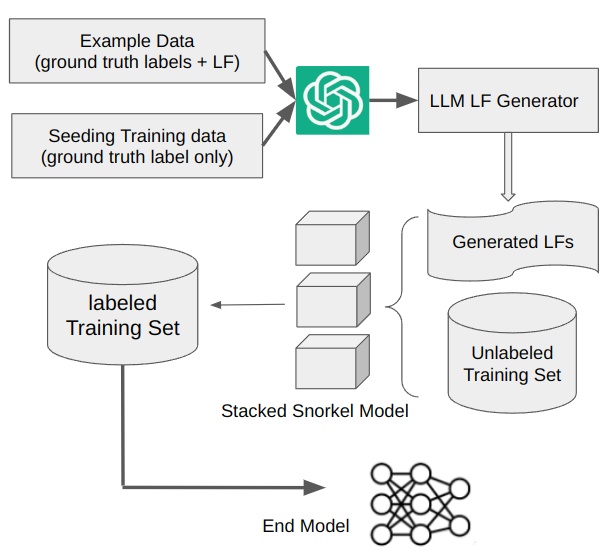}
	% \vspace{-1em}
	\caption{Overall Workflow of Labeling Pipeline with LLM}
	\label{fig:system}
\end{figure}

The overall architecture of our proposed pipeline is shown in Figure~\ref{fig:system}.
Similar to previous weak supervision works, it starts with some seed instances that help provide signals to generate LFs. 
Then we ask LLM to generate LFs based on the provided seed instances.
To this end, we conduct few-shot learning by providing some examples of the pairs of instances and LFs generated from them.
After obtaining the set of LFs, we use them along with the seed instances to train the label model and obtain the aggregated results.
In the current pipeline, we choose the well-known Snorkel~\cite{DBLP:journals/pvldb/RatnerBEFWR17} framework as the label model.
Then given an unlabeled instance, i.e. column from a table, we will feed it to the label model and obtain the labels.
Here we could evaluate the quality of the label model by considering the accuracy of labeling a set of unlabeled instances.
Finally, we regard the instances obtained from the label model as the training set for an end model.

\subsection{Labeling Function Generation with LLM}\label{subsec-prompt}
% \begin{figure}[ht]
% 	\centering
% 	\includegraphics[width=0.5\textwidth]{fig/prompt.png}
% 	\vspace{-2em}
% 	\caption{The Prompt Template for LF Generation}
% 	\label{fig:prompt}
% \end{figure}
\begin{figure*}[ht]
	\centering
	\includegraphics[width=
 \textwidth]{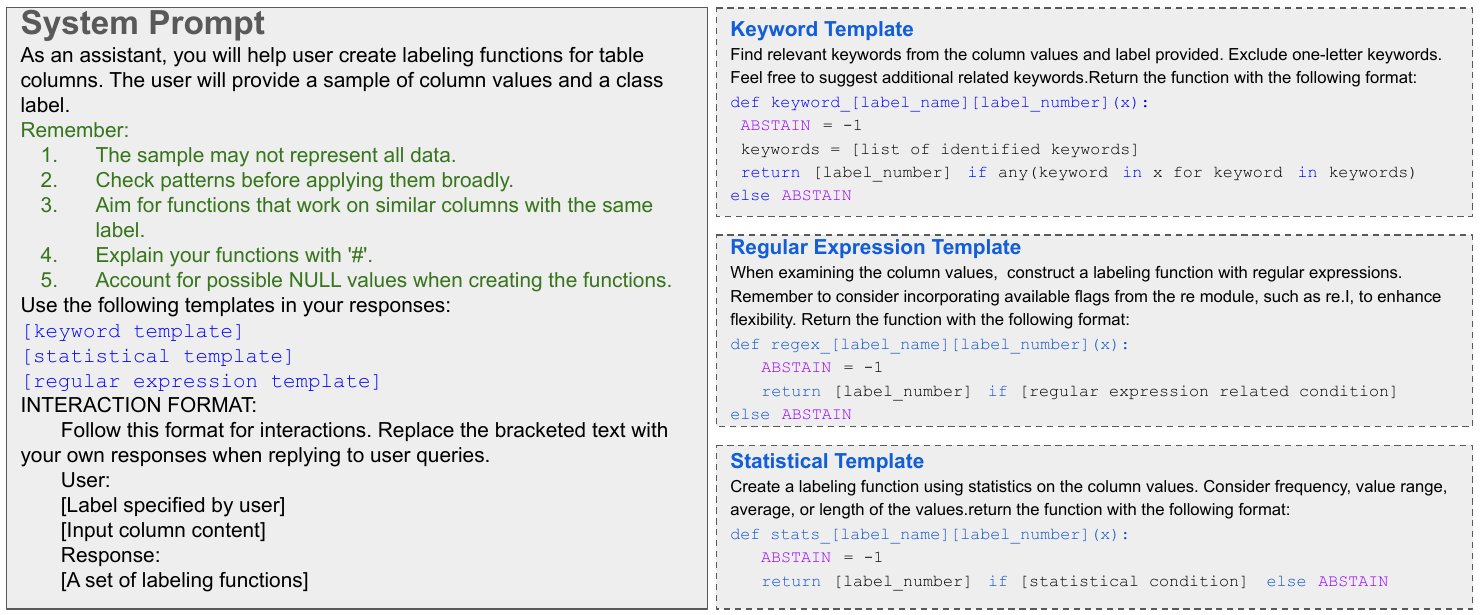}
	%\vspace{-2em}
	\caption{The Prompt Template for LF Generation}
	\label{fig:prompt}
\end{figure*}

Next we introduce how to generate LFs by leveraging the LLMs.
We follow previous studies~\cite{DBLP:journals/pvldb/RatnerBEFWR17,DBLP:journals/pvldb/HsiehZR22} to select the LFs based on the following aspects:
\begin{itemize}
	\item Keyword: This kind of LF assumes that for each given semantic type $L$, there is a list of common keywords. Once the overlap between the unlabeled column and the list reaches a certain threshold, the column should be given the label $L$.
	\item Statistical: This kind of LF decides the label of a column based on the pattern of value distributions. The label is assigned if the value distribution of a column satisfies some pre-defined statistical patterns.
	\item Regular Expression: The LF could also be expressed with regular expressions. The label will be assigned if the column values match a provided regular expression.
\end{itemize}
%In this work, we will generate all three kinds of LFs for each semantic type. 
%\chenjie{this is not entirely true as some semantic type may not get certain kind, we did not require each semantic type will get all three kinds}

% para 2: introduce the options for prompt and our prompt template
The prompt template for generating is shown in Figure~\ref{fig:prompt}.
It consists of two components: the system prompt and the user prompt.
The system prompt is the general description of the background as well as some necessary instructions for the LLM.
For example, our task of semantic type detection needs to specify the input as the given column values and the output as the semantic type of the column.
The user prompt aims to provide the necessary contextual information to generate the labeling functions.
In our prompt template, we employ the few-shot learning approach and provide some selected examples as demonstrations with the following information: (i) the examples with pairs of column values and ground truth semantic type; (ii) the LF template where the three kinds of LFs introduced above will have different templates; (iii) The question to ask for labeling functions. 
We randomly selected 5 semantic types and manually wrote a few labeling functions with the 3 kinds of LFs mentioned before. Specifically, within each group, we randomly selected 5 cell values delimited by a special character from that column to be used in the demonstration. 
%The demonstration could also be selected in different ways according to previous studies about general demonstration selection methods~\cite{DBLP:conf/nips/BrownMRSKDNSSAA20} as well as those specifically defined for weak supervision tasks~\cite{DBLP:journals/pvldb/HsiehZR22}.
%Due to the space limitation, we leave it as future work.

% para 3: new strategy: include ground truth 
Compared with previous studies that try to use LLM to generate LFs ~\cite{DBLP:conf/acl/ZhangYSSZ22,DBLP:journals/corr/abs-2311-00739}, we improve in the process of prompt construction.
Specifically, we include the ground truth of the semantic type label of the given column in the prompt in the above item (i).
The reason is that since the task is to generate the labeling function instead of predicting the label, it does not result in the risk of ground truth leakage.
Meanwhile, providing the ground truth label could help LLM obtain more information to produce the labeling function as it does not need an additional step to predict the label.
%Besides, it could also eliminate the potential noises caused by the mistakes in predicting ground truth labels.
% \vspace{-3.5mm}
\subsection{Stacked Labeling Model}\label{subsec-stacked}

% para 1: the motivation of performing stack, complexity of Snorkel label model
With the LFs generated by LLMs, the next step is to filter out the LFs with low quality.
We use the idea of accuracy and redundancy filter introduced in the previous study~\cite{DBLP:journals/corr/abs-2311-00739}.
After that, we fit the remaining LFs into the Snorkel label model and aggregate the labeling results.
To reach this goal, there is still another challenge to overcome: the space complexity of Snorkel is $\mathcal{O} (N*M*d)$, where $N$ is the number of seed instances, $M$ is the number of semantic types, $d$ is the number of remaining LFs.
Compared with those in previous studies of data programming~\cite{DBLP:conf/nips/ZhangYNWYYR21}, the target label set of semantic type detection is up to an order of magnitude larger. 
As a result, there will also be a much larger number of LFs which will bring significant overhead to the label model.

% para 2: our practice and potential design space
In this work, we use a stacking-based solution to solve this problem.
The basic idea is to split the set of semantic type labels into several disjoint groups and train a Snorkel label model for each group.
Then the overall computation cost will be significantly reduced.
We will also have a routing model stacked on top of the set of label models: when a new unlabeled instance arrives, it sends it to all the label models and decides the label based on the results of all label models.
In the current implementation, we choose the label with the highest probability as the result.
The next issue to be resolved is how to split the set of labels.
Some datasets, such as TURL WikiTables~\cite{DBLP:journals/pvldb/DengSL0020}, provide the hierarchical structure of the label set which can be directly utilized to create groups.
For other general cases, we propose to first get the word embedding of each label and then perform a K-means clustering over the word embeddings, where $K$ target group count.

%% file: src/sec3-exp.tex
\section{Experiments}\label{sec-exp}

\subsection{Experiment Setup}\label{subsec-setup}

We evaluate the proposed framework on widely-used benchmark for semantic type detection. The Viznet dataset~\footnote{https://github.com/megagonlabs/sato/tree/master/table\_data} is processed in the previous study~\cite{DBLP:journals/pvldb/ZhangSLHDT20} on the basis of the  Viznet corpus.
There are 78 column types and 119,360 columns from 78,733 tables in total.
 We also explore a more challenging WikiTable dataset \footnote{https://github.com/sunlab-osu/TURL\#data} proposed in~\cite{DBLP:journals/pvldb/DengSL0020}, consisting of 570,171 tables with 255 semantic types.   
%The ground truth of semantic type detection is obtained by aligning Freebase with the web tables.
%There are 255 semantic types and 570,171 tables in total.
%The training set consists of 628,254 columns from 397,098 tables, while the validation and test set consists of 13,391 columns from 4,844 tables and 13,025 columns from 4,764 tables, respectively.
%Specifically, we select $32226$ and $95540$ unlabeled instances from WikiTable and Viznet, respectively.

% para 2: evaluation metrics, implementation, baseline 
To construct prompt for labeling function generation, for each semantic type in each dataset, we randomly select 10 columns that share the same semantic type and randomly select 5 column values from each column as the seed instance. We use GPT-4 as the LLM for labeling function generation.
With the set of labeling functions, we stack 5 smaller snorkel models to generate the augmented training set. Finally, we fine-tune DoSolo, the single task version of Doduo~\cite{DBLP:conf/sigmod/SuharaL0ZDCT22}  as the end model.
For the $F_1$ score, we report results of both Micro and Macro $F_1$ scores. 

% We implement the pipeline in Python using Pytorch and the Hugging Face Transformers library.
%\dz{CRF setup}
%Then we randomly select some instances that are not overlapped with the seed instance as the unlabeled instances and obtain the training set with the help of the label model.

\subsection{Results}\label{subsec-result}

\begin{table}[ht]
	\centering
	\caption{Step by step evaluation results for the Viznet dataset.}\label{tbl-res}
	\vspace{-1em}
	\begin{tabular}{l|ccc}
	\toprule
	Evaluation Step & Metric  & Value(\%) \\
	\midrule
        Labeling Function & Avg. F1 & 18 \\
	\midrule
	\multirow{2}{*}{Label Model} & Micro $F_1$ & 22  \\ 
	& Macro $F_1$ & 28  \\ 
	\midrule
	\multirow{2}{*}{End Model} 
	& Micro $F_1$ & 43 \\ 
	& Macro $F_1$ & 31 \\ 
	\bottomrule
	\end{tabular}
\end{table}

% \begin{table}[ht]
% 	\centering
% 	\caption{Main Results}\label{tbl-res}
% 	\vspace{-1em}
% 	\begin{tabular}{l|ccc}
% 	\toprule
% 	Dataset& Metric & WikiTable & Viznet \\
% 	\midrule
% 	\multirow{3}{*}{Labeling Function} & Accuracy & \valph & \valph \\ 
% 	& Micro $F_1$ & \valph & \valph \\ 
% 	& Macro $F_1$ & \valph & \valph \\ 
% 	\midrule
% 	\multirow{3}{*}{End Model} & Accuracy & \valph & \valph \\ 
% 	& Micro $F_1$ & 6.5 & 43 \\ 
% 	& Macro $F_1$ & 9.4 & 31 \\ 
% 	\bottomrule
% 	\end{tabular}
% \end{table}
%To evaluate the performance of our proposed pipeline, we look at metrics from two aspects: the performance of label model and results of training the end model with the created dataset.

The step-by-step evaluation results for the Viznet data obtained for the proposed pipeline are shown in Table~\ref{tbl-res}. We directly evaluate the quality of the LLM-generated labeling functions, the output of snorkel inference and final prediction of the end model fine-tuned on the noisy labels obtained from our augmentation process. 
Although there is a gap in performance between the end model and ones from previous studies that performs fine-tuning over pre-trained language models~\cite{DBLP:conf/sigmod/SuharaL0ZDCT22,DBLP:journals/pacmmod/MiaoW23,DBLP:journals/pvldb/DengSL0020}, it is worth noting that we only use a very limited set of labeled instances, which is up to 1.3\% of the full training set as demonstration for the LLMs. %We need to obtain additional training signals from the label model while previous studies train the same end model using the whole labeled training set.
We also evaluated the pipeline on a more challenging Wikitables dataset with 255 candidate classes and the fine-tuned end model has micro-F1 of 0.065 and macro-F1 of 0.094.

Compared to popular tasks (e.g., binary classification, NLI) in previous data programming works~\cite{DBLP:conf/nips/ZhangYNWYYR21}, semantic type detection tasks have much more complicated label spaces.
Therefore, the task of finding explicit labeling functions (in the format of three kinds introduced before) for semantic type detection itself is rather challenging.
%Due to the limited expressive power of explicit labeling functions, it is difficult to find a small set of labeling functions with both good coverage and effectiveness. Since we only have 800 labeling functions to train the Snorkel label model, the overall performance of labeling functions is not so promising as shown in Table~\ref{tbl-res}. As a result, it will also not be able to help generate a high-quality training set for the end model.
One takeaway from our initial results would be that while explicit labeling functions works well for simple tasks such as text classification introduced in previous studies of data programming~\cite{DBLP:conf/nips/ZhangYNWYYR21,DBLP:conf/nips/RatnerSWSR16,DBLP:journals/pvldb/RatnerBEFWR17}, for more complicated tasks like semantic type detection, we might either need to collect a much larger number of labeling functions and developing more scalable and efficient methods to train the label model; or develop new kinds of labeling functions that can trade the transparency for effectiveness.

\subsection{Case Study}\label{subsec-case}

% \begin{figure*}[ht]
% 	\centering
% 	\includegraphics[width=0.8\textwidth]{fig/lfexamples}
% 	\vspace{-1em}
% 	\caption{Examples of Generated Labeling Functions}
% 	\label{fig:lf}
% \end{figure*}

\begin{figure}[ht]
    \centering
    \begin{subfigure}[b]{0.5\textwidth} % Reduced width
        \centering
        \includegraphics[width=\textwidth]{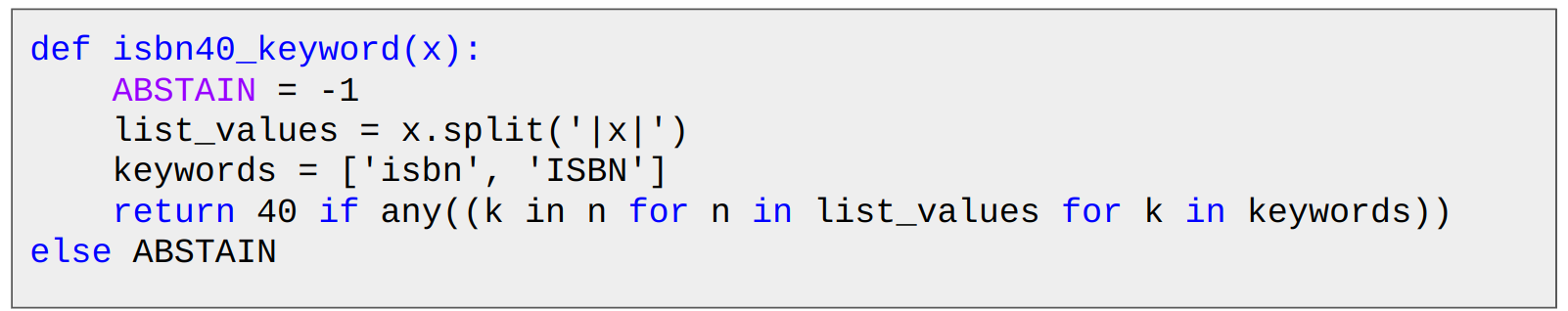}
        \caption{Keyword LF for semantic type ISBN}
        \label{fig:keyword_example}
    \end{subfigure}
    \hfill % Adjust or remove if necessary
    \begin{subfigure}[b]{0.5\textwidth} % Reduced width
        \centering
        \includegraphics[width=\textwidth]{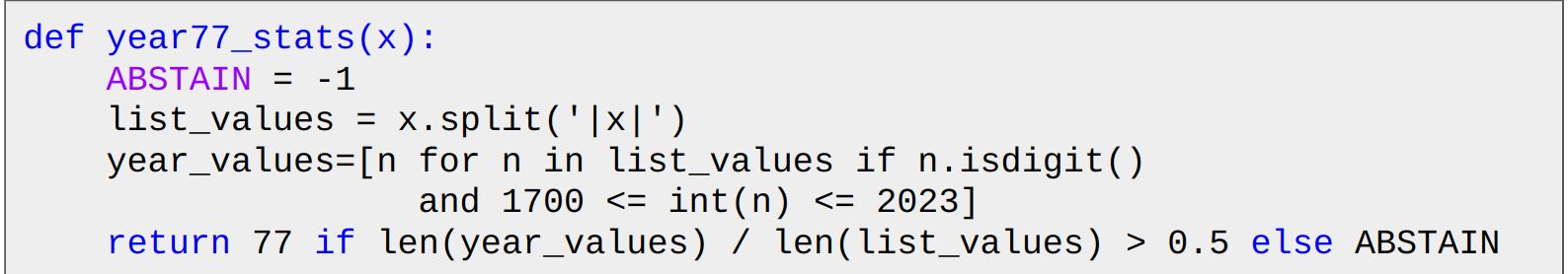}
        \caption{Statistics LF for semantic type year}
        \label{fig:stats_example}
    \end{subfigure}
    \vskip\baselineskip
    \begin{subfigure}[b]{0.5\textwidth} % Reduced width
        \centering
        \includegraphics[width=\textwidth]{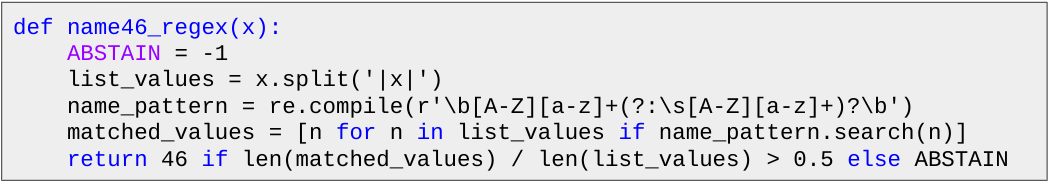}
        \caption{Regular Expression LF for semantic type name}
        \label{fig:too_general_lf}
    \end{subfigure}
    \caption{Examples of Generated Labeling Functions}
    \label{lf_examples}
\end{figure}

Finally, we conduct a case study to show some specific labeling functions generated by our proposed pipeline in Figure~\ref{lf_examples}.
Generally, we observe that the LFs generated by LLM could express reasonable semantics and are comparable with those written by humans.
For example, Figure~\ref{fig:keyword_example} illustrates a keyword-based LF for semantic type ISBN from Viznet dataset. As this column has a simple format (most of the column values are ISBN followed by the product identification number), it provides a very accurate rule to identify columns that belong to the ISBN type.
Figure~\ref{fig:stats_example} is a statistic-based LF for semantic type year. As shown in the function, LLM could incorporate its internal knowledge with the given semantic type. Specifically, in the return statement it added the frequently mentioned year range (1700-2023) as the way to decide if a string is a year or not.
% This is recognized by the LLM from the input prompt about the  \red{name} type.
Meanwhile, there are also some examples that LF has good coverage but is not so accurate.
The LF in  Figure~\ref{fig:too_general_lf}  is a function created for the semantic type ``name''. The function looks for words that start with a capital letter followed by lowercase letters, optionally followed by another similar word. This function can cover most of the names, but at the same time, it can also cover a lot of values that are from the other semantic types such as location or address. Thus it provides limited insights in labeling the instances.

% means that once the column satisfied the regular expression, it should belong to the \emph{location} type. 
% However, such a LF is too general and the semantic type detection task needs to identify more accurate kind of location, such as country, city and street etc.

%% file: src/sec4-conclusion.tex
\section{Conclusion and Future Work}\label{sec-conc}

In this paper, we studied the problem of generating labeling functions for the task of semantic type detection.
We proposed an end-to-end pipeline that adopted LLM to automatically generate labeling functions for semantic type detection via prompt engineering and train a label model based on Snorkel.
% To cope with the challenges of large label space and a huge number of labeling functions, we propose a stacked label model that can significantly improve the scalability while keeping a reasonable level of effectiveness.
We make an extensive set of explorations on the design space and conduct initial experiments on two popular benchmark datasets.

Based on our initial efforts, we recognize several essential directions for further exploration of this topic. As illustrated by our experimental results, the current explicit labeling functions commonly used in previous data programming studies may not be sufficient for handling more complex tasks like semantic type detection. For this task, as well as related table understanding tasks such as relationship extraction and column population, we need to develop new types of labeling functions that are more powerful yet still explainable. A promising starting point could be building labeling functions based on simple machine learning models, such as logistic regression and decision trees. It is also crucial to improve the scalability of the label models concerning the number of class labels and labeling functions. Our idea of splitting the label space provides a reasonable solution to this problem. It is necessary to explore this approach further by generalizing the problem and developing an efficient algorithm that can identify high-quality splits.

Moreover, it is beneficial to consider advanced label models developed in recent efforts from the machine learning community as introduced in~\cite{DBLP:journals/corr/abs-2202-05433}.
Last but not least, although the effectiveness of labeling functions generated in this work is limited, the labeling functions could still provide some useful insights for the semantic labels of columns. 
So it is also worth investigating how to use the generated labeling function to explain the results of existing solutions for semantic type detection, such as those based on pre-trained language models. 
% In this way, labeling functions could help improve the transparency of previous studies in this field.

% selection of demonstration
%The demonstration could also be selected in different ways according to previous studies about general demonstration selection methods~\cite{DBLP:conf/nips/BrownMRSKDNSSAA20} as well as those specifically defined for weak supervision tasks~\cite{DBLP:journals/pvldb/HsiehZR22}.